# A multiscale model for the slow viscoelastic response of liquid foams

Sébastien Vincent-Bonnieu, Reinhard Höhler, Sylvie Cohen-Addad

Université de Marne-la-Vallée,

Laboratoire de Physique des Matériaux Divisés et des Interfaces, UMR CNRS 8108

5 Boulevard Descartes, 77 454 Marne-la-Vallée cedex 2, France

Using 2D numerical simulations as well as analytical modelling, we show how slow viscoelastic dynamics of aqueous foam are linked to coarsening induced intermittent bubble rearrangements. The macroscopic strain rate is expressed in terms of mesoscopic dynamics where local rearrangements are represented by strain fields induced by force dipoles. We relate these dipoles quantitatively to local geometrical features of the corresponding bubble rearrangements. We also explain how applying a small macroscopic stress biases the outcome of coarsening induced rearrangements.

To flow on a macroscopic scale, liquid foams[1], emulsions, pastes, amorphous metals and other disordered close packings of small units must undergo irreversible rearrangements on the scale of the bubbles, droplets, grains or atoms. These dynamics may be driven directly by the applied stress if it is strong enough to unjam the structure and to induce plastic flow. Stresses below a limit called yield stress don't disrupt the packing but they can induce viscoelastic creep flow, related to rearrangements induced by thermal fluctuations or other intrinsic dynamics. Additional frequently observed common features of the mentioned class of materials are slow relaxations coupled to ageing, reminiscent of glassy behaviour[2], and mechanical memory effects[3]. In the recent literature, "Soft glassy rheology"[2] or "Shear transformation zone" [4, 5] mean field models have been proposed that relate macroscopic flow to local rearrangement dynamics. They describe flow induced non-thermal fluctuations by an effective temperature[6, 7] and they are formulated on a mesoscopic length scale, sufficiently large so that complex details of the microstructure can be ignored but sufficiently small so that spatial variations of mechanical properties within the sample are resolved. Two crucial questions arise in this context: Which minimal set of parameters characterizing a rearrangement mesoscopically is required so that its contribution to the macroscopic flow can be predicted?  How are these mesoscopic parameters linked quantitatively to the microstructure of a given material and how are they coupled to the applied stress? In this paper, we provide an answer to the first question that we expect to be useful for a wide class of materials, ranging from foams to amorphous metals.  The second question is studied numerically for 2D liquid foams, a model system that many previous authors have investigated numerically and experimentally, as discussed in recent reviews[1, 8]. We focus on the slow viscoelastic creep flow of these materials and report laws and mechanisms linking macroscopic flow, mesoscopic dynamics and microstructure changes in the quastistatic case. The results are compared to previous 3D experimental data and models.

Let us consider an elastic incompressible material in which a local structural rearrangement event occurs at a position **x'**, inducing displacements **u(x-x')** at neighbouring positions **x**. Our aim is to understand how such an event can contribute to macroscopic flow by generalizing previous results [9-11]. Since only the behaviour of **u(x-x')** far from **x'** is relevant here, **u(x-x')** can be simplified using the leading order term of a multipole expansion [10]. This term is dipolar and we represent the displacement induced far from **x'** as the effect of two fictive forces, **-F** and **F,** acting at positions **x' - d**/2 and **x' + d**/2. The dipolar far field of **u(x-x')** can be related as follows to **F** and **d** using the Green's function of Navier's equation, denoted as $\Gamma_{ik}$ [9, 10]:

$$u_i(\mathbf{x}-\mathbf{x'}) = -P_{sk}\frac{\partial \Gamma_{ik}(x_s - x'_s)}{\partial x_s} \qquad (1)$$

The dipole tensor $P_{sk}$, defined as $P_{sk} = F_s d_k$, has been used previously in the context of point defects in metals[10] and dilute emulsion rheology[9]. $P_{sk}$ is fully specified by its eigenvalues, denoted as $p_1$ and $p_2$, and the angle $\alpha$ of the rotation leading from the $x_1$ axis of the laboratory frame to the frame where $P_{sk}$ is diagonal. Physically, $\alpha$ describes an intrinsic direction associated to the event whereas $p_1$ and $p_2$ are measures of the dipole strength. To model the flow of 2D foams and similar materials, we now focus on the effect of rearrangements with given $\alpha$, $p_1$ and $p_2$, occurring at random locations, with a concentration per area denoted as $\rho$. **U(x)** is the expected displacement field, obtained by averaging over all possible event positions **x'**:

$$\mathbf{U}(\mathbf{x}) = \rho \int_{-\infty}^{\infty}\int_{-\infty}^{\infty} \mathbf{u}(\mathbf{x}-\mathbf{x'})\, dx'_1\, dx'_2 \qquad (2)$$

To predict the rearrangement induced macroscopic shear strain $\delta\gamma$ in the $x_2$ direction, we calculate the variation of $U_2(\mathbf{x})$ per unit length in the $x_1$ direction: Eq.(1) is substituted into Eq.(2) and $P_{ij}$ is expressed in terms of $p_1$, $p_2$ and $\alpha$. By evaluating the integral in Eq. 2 using the expression of $\Gamma_{ik}$ given in the literature[12] we then obtain:

$$\delta\gamma = \frac{p_1 - p_2}{2G}\rho \sin(2\alpha) \qquad (3)$$

This general prediction of the macroscopic shear contains a minimal set of parameters that provides all necessary information about the randomly placed rearrangements: Besides the scalar measure of event strength $p_1$-$p_2$ the event orientation, described by the angle $\alpha$, must be specified. Indeed, numerical simulations of 2D foams undergoing plastic flow have provided first qualitative evidence of anisotropic, multipolar stress variation patterns in the vicinity of rearrangements[13]. The importance of event orientation has been recognized in the framework of STZ models where this feature has been implemented in 1D[5]. Let us point out that Eq.(3) is consistent with a previous analytical study where rearrangement events represented by two pairs of forces inducing simple shear but no torque were studied in detail in the context of plastic flow[11]. However, in this latter work only forces of fixed orientation were considered, corresponding in our notation to $\alpha = \pi/4$.

We will now discuss an example where the intrinsic direction of rearrangements is of crucial importance: The linear viscoelastic creep flow of foams. This phenomenon and the closely related low frequency oscillatory rheological response have recently been studied experimentally[14-16], numerically[14, 15, 17] and theoretically[2]. Here, we focus on the case where the applied stress is smaller than the yield stress, and where rearrangements can only be triggered by an aging process called coarsening: Driven by Laplace pressure differences, gas diffuses through the liquid films separating neighbouring bubbles, giving intermittently rise to unstable bubble configurations that undergo neighbour switching rearrangements called T1 events[8]. Examples are shown in Figure 1. These intrinsic dynamics progressively unjam the foam structure[15] so that it flows in response to an applied shear stress $\sigma$ at a macroscopic creep flow rate $\dot{\gamma}$[14]. In recent 3D rheological experiments, the rate of T1s per unit time and volume, denoted as R, was detected in situ using a light scattering probe. The data were found to scale as follows[14]:

$$\dot{\gamma} \approx \frac{\sigma R V}{G} \qquad (4)$$

$G$ is the shear modulus and $V$ an empirical parameter. In 2D foams, $V$ has a dimension of area and $R$ is a rate per unit time and area. Data consistent with Eq. (4) have also been obtained in recent quasistatic 2D numerical creep simulations of coarsening foams[17] which demonstrate that macroscopic creep flow proceeds by discrete steps that systematically coincide with T1 events. These experiments and simulations were interpreted using a simple mesoscopic continuum mechanics argument, explaining creep flow as the progressive accumulation of strain steps due transient stress relaxations upon each T1[14, 17]. In this framework, the parameter V in Eq.(4) was interpreted as an effective volume (or area in 2D) where stress is locally relaxed upon a T1. However, applying continuum mechanics on the bubble scale is a rough approximation. In this Letter, we therefore report and analyse simulations of creep flow in coarsening 2D foams that reveal the mechanism of flow on mesoscopic and local length scales. 10 samples of polydisperse 2D foam of vanishing liquid content with 400 bubbles each are created in an initially rectangular unit cell, using a Voronoi network based on randomly dispersed points (for an illustration of similar structures, see[1]). Since in the 3D creep experiments[14], wall effects were found to be negligible we use periodic boundary conditions in the simulations to exclude such effects. Using the Surface Evolver software[18] equilibrium foam structures of minimal interfacial energy are determined. The interfacial stress is calculated using Eq.(5) [19, 20] where **t** is the unit vector tangent to a bubble edge, and T the edge line tension. The integration is carried out over all the edges of the sample unit cell whose area is denoted as A :

$$\sigma_{ij} = \frac{T}{A} \int t_i t_j \, dl \qquad (5)$$

We perform quasistatic creep simulations as follows[17]: the periodic boundary conditions defining the average strain in the $x_2$ direction are continually adjusted such that the shear stress obtained using Eq.(5) keeps a constant value. To adjust strain, we first impose affine shear and then relax the structure so that the interfacial energy is minimal. At the same time, coarsening is implemented as reported in[17]. T1s and the macroscopic strain steps δγ that they induce are systematically monitored. Strain vs time curves as well as detailed strain step statistics are reported in[17]. Here, we go further and investigate how the parameters $p_1$, $p_2$ and α defining the dipole tensor are linked to the local microstructure change. Upon each T1, a new bubble edge is created that naturally defines a characteristic direction (cf. Fig 1) and we therefore conjecture that it sets the value of α. For dimensional reasons, the dipole strength of the T1 must be the product of a force and a length, suggesting that it should scale as the edge line tension $T$ times the length of the new edge $\lambda$. By combining these dimensional and symmetry arguments with Eq.(3), we predict the average strain step $\delta\gamma$ due to a single T1 in the sample:

$$\delta\gamma \propto T\,\lambda\,\sin(2\alpha)/(G\,A), \qquad (6)$$

We validate Eq.(6) and determine the dimensionless prefactor using our simulations: Figure 2a) shows the predicted sinusoidal evolution of $\delta\gamma$ with $\alpha$ that is independent of applied stress. Moreover, Figure 3 confirms that $\delta\gamma$ scales linearly with $\lambda$ as well as $\lambda\sin(2\alpha)$, again in full agreement with Eq.(6). Note that the shear strain steps $\delta\gamma$ upon T1s shown on Fig. 2 are associated to new bubble edges whose angles $\alpha$ are of the same sign as $\delta\gamma$, implying that the foam is stretched in the direction of the new edge (cf. Fig 1). This is remarkable, since the contribution to the macroscopic stress due to this edge is a tension in the direction tangential to it (cf. Eq. (5)). If this contribution were dominant, one would expect a strain opposite to the observed stretching. Therefore, even though key parameters can be extracted from the edge created due to a T1, this edge itself provides only a minor contribution to the total stress evolution upon a T1. The entire complex microstructure change in the neighbourhood of a T1 must be taken into account to obtain this latter quantity.

To understand creep flow, the link between microstructure changes and macroscopic shear that we have established is not enough; we must also explain how the orientations of the edges created upon T1s are related to the applied stress. Since this stress is too weak to induce by itself any T1s, all it can do is bias the outcome of coarsening induced T1s. Let us revisit

Figure 1 which illustrates schematically what happens when a small bubble looses its gas due to coarsening. Small bubbles tend to have few neighbours[1], and we find that 85% of all the observed T1's are related to 4 sided bubbles that become 3 sided ones. Depending on which of the 4 edges first reaches zero length this process can lead to one of two possible orientations of the new edge, corresponding to different macroscopic strain steps (cf. Figure 1). The selection among them can be biased by applying prior to the T1 a macroscopic stress that distorts the small bubble. We evaluate the integral in Eq. (5) only over its edges to measure this distortion. The resulting tensor is the bubble's contribution to the macroscopic stress. The eigenvector associated to the largest eigenvalue of this stress contribution indicates the dominant direction of bubble elongation. Let us call $\alpha_B$ the angle between this direction and the $x_1$ axis. Figure 4a demonstrates a strong linear correlation between $\alpha_B$ and $\alpha$, for all T1s in a coarsening foam, confirming that local stress determines which final microstructure is obtained after a T1. To investigate the consequences of this feature, we study the probabilities of finding $\alpha$ in the intervals $[-\pi/2, 0]$ and $[0, \pi/2]$, respectively denoted as $P_-$ and $P_+$, as a function of applied stress $\sigma$. Figure 2b shows that $P_+$ and $P_-$ are equal only for $\sigma = 0$, reflecting the statistical isotropy of the foam in that case. Our data for stresses in the range $-0.25\ \sigma_y < \sigma < 0.25\ \sigma_y$ shown in Figure 4b are well described by the empirical law $P_+ - P_- = \chi\sigma$. Combining all of our findings allows the macroscopic creep rate to be expressed as:

$$\dot{\gamma} = \frac{2RA}{\pi}\left( P_- \int_{-\pi/2}^{0} \langle\delta\gamma(\alpha)\rangle d\alpha + P_+ \int_{0}^{+\pi/2} \langle\delta\gamma(\alpha)\rangle d\alpha \right) \approx \frac{\sigma R \chi T \langle\lambda\rangle}{G} \qquad (7)$$

The average represented by the symbol $\langle\ \rangle$ is performed over the distribution of new edge lengths $\lambda$. Remarkably, Eq (7) agrees with the scaling behaviour Eq. (4) observed in 3D experiments[14]. Comparing these two equations reveals the physical meaning of the phenomenological parameter V in terms of the elastic dipole strength $T \langle\lambda\rangle$ of a T1 and the susceptibility $\chi$. Moreover, several of our simulation results remind features of STZ theory which was developed on the basis of molecular dynamics simulations of amorphous metals[4]: In this context as well as for the foams studied here, flow is governed by "defects" behaving as two state systems coupled to the applied stress. There may therefore indeed be generic concepts useful for describing materials as different as dry 2D foams and amorphous metals, provided the specific origin of rearrangement dynamics is taken into account. As a perspective, we point out that Eq. (6) also applies to individual T1s occurring in plastic flow of 2D foams as we will show and discuss elsewhere.

To conclude, we have shown that the link between macroscopic flow and local bubble rearrangements in dry 2D foam undergoing slow viscoelastic creep flow can be predicted by a mesoscopic model where the rearrangements, driven by coarsening, are represented by force dipoles. We go beyond similar previous models by showing how the dipole strength and direction are quantitatively related to changes of the microstructure upon a rearrangement whose relevant features are the length and orientation of the bubble edge created upon a T1. In this framework, creep flow is explained as the consequence of a stress induced bias of coarsening induced bubble rearrangements, in contrast with previous interpretations in terms of soft glassy dynamics. These results open the perspective of a physical multiscale model without phenomenological parameters explaining slow viscoelastic dynamics and plastic flow in foams and, in a wider context, a generic framework for rearrangement driven flow that could apply to a large class of disordered materials.

Acknowledgements - We thank M. Dennin and the participants of the Foam Rheology in Two Dimensions Workshop held in Aberystwyth in 2005 for stimulating discussions.

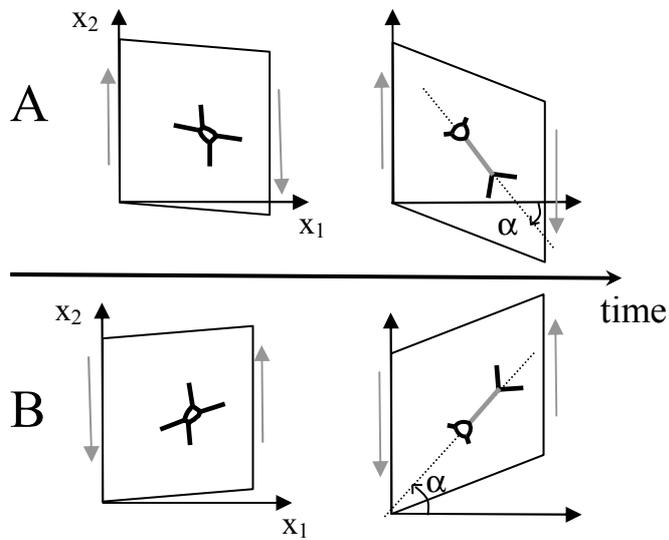

**Figure 1** : Schematic view of a given 4 sided bubble in a 2D foam whose shape differs for two applied shear stresses σ, symbolized by the grey arrows in parts A) and B). The bubble shrinks due to coarsening until the shortest of its edges vanishes, and a new edge highlighted in gray is created upon a T1 event, as shown on the right of A) and B). The T1 is also accompanied by a macroscopic strain step δγ, illustrated by the shear of the bounding box. A) and B) show that depending on σ two final structures are possible, characterized by different angles α between the new edge and the $x_1$ axis. Note that upon the T1, the sample is stretched in the direction of the new edge.

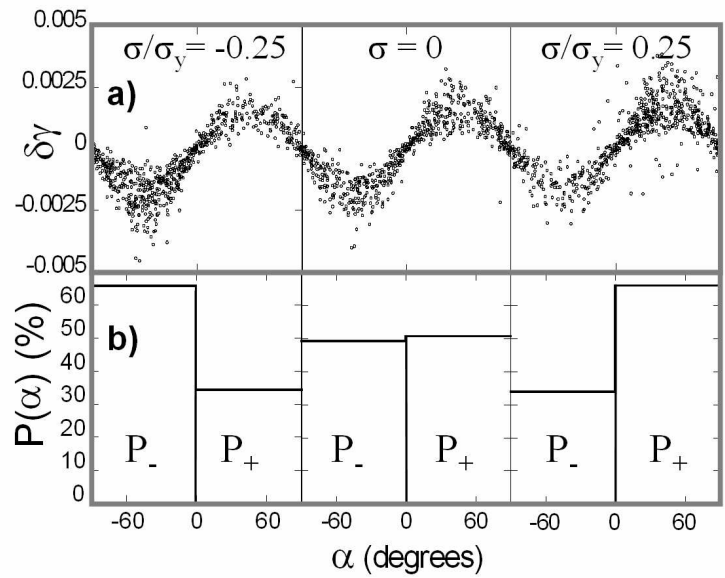

**Figure 2 :** a) Macroscopic strain step $\delta\gamma$ due to an individual coarsening induced T1 versus angle $\alpha$ of the edge created upon the rearrangement. Data are shown for 3 different applied stresses $\sigma$ normalized by the yield stress $\sigma_y$. b) Probability of finding edges with positive or negative $\alpha$, for the 3 stresses shown in a).

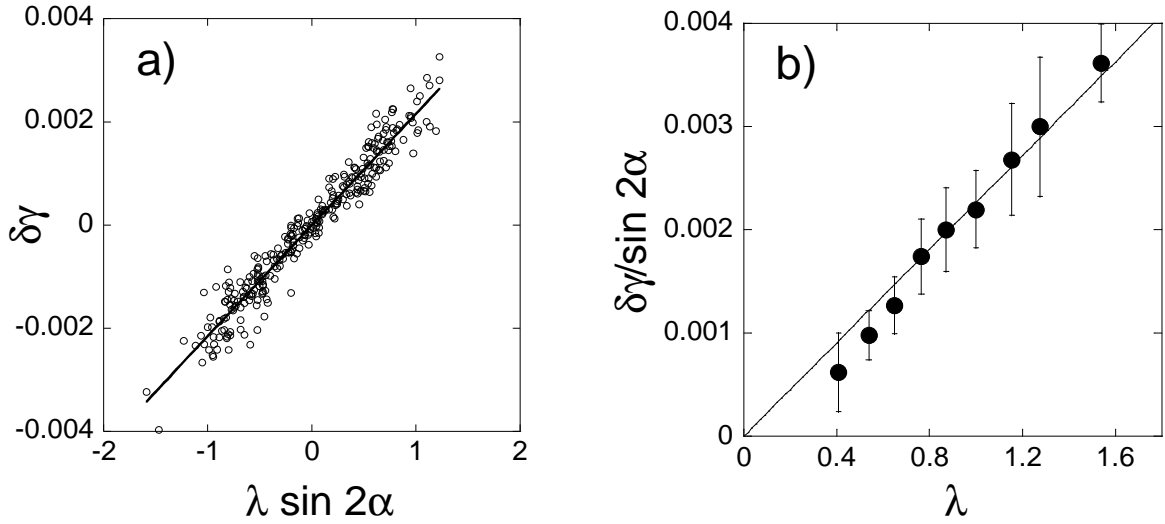

**Figure 3**: a) Strain steps $\delta\gamma$ due to all the T1s shown in figure 2a) versus $\lambda \sin(2\alpha)$. The line is a linear fit with a slope $0.0021 \pm 2\ 10^{-5}$ b) Strain steps due to T1s, divided by $\sin(2\alpha)$, binned and plotted as a function of $\lambda$. The line is a linear fit with a slope of $0.0022 \pm 6\ 10^{-5}$. In a) and b) $\lambda$ is expressed in units of the mean bubble radius at the beginning of the creep experiment.

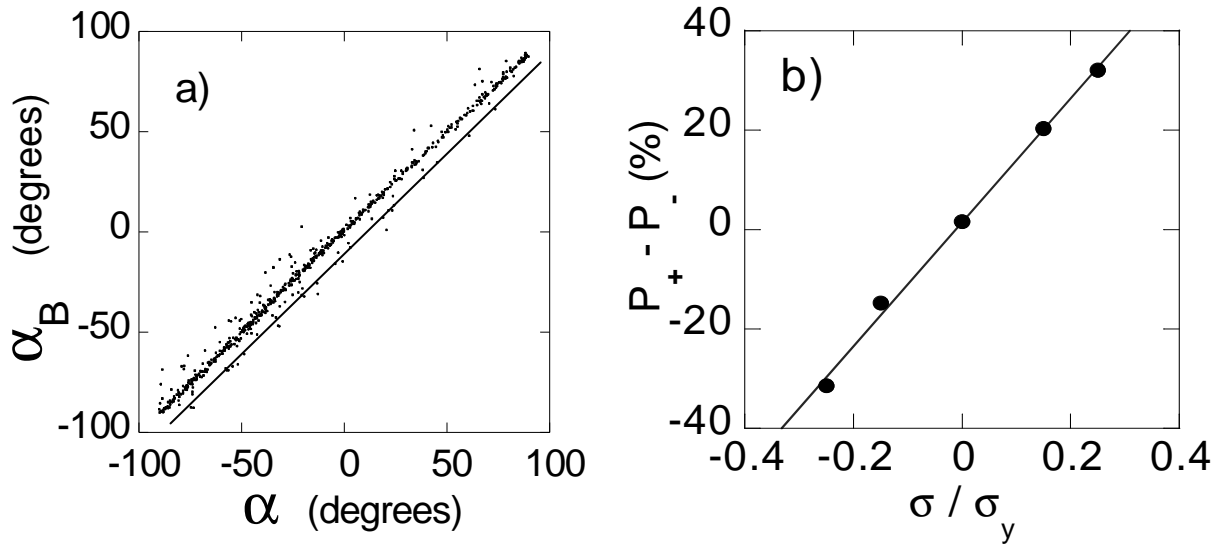

**Figure 4**: a) $\alpha_B$ describes the direction of elongation of a small bubble about to undergo a T1. $\alpha$ describes the direction of the edge created upon this T1. The line is a linear fit of slope 0.994 ± 0.003, translated by 10° so that the data points are not hidden. b) The imbalance $P_+-P_-$ defined in the text, versus the applied stress, scaled by the yield stress. The line is a linear fit with a slope 125 ± 6.